\documentclass[aps,prc,showpacs,twocolumn,nofootinbib,preprintnumbers]{revtex4-1}

\usepackage{longtable}
\usepackage{graphicx}
\usepackage{dcolumn}
\usepackage{bm}
\usepackage{amsmath,amssymb}
\usepackage{epstopdf}
\usepackage{hyperref}

\DeclareMathOperator\artanh{artanh}

\usepackage[normalem]{ulem}  
\usepackage[dvips]{color} 

\renewcommand\sout{\bgroup \color{red} \ULdepth=-.5ex \ULset}

\begin{document}

\preprint{YITP-16-83}


\title{Quark mass dependence of H-dibaryon in $\Lambda\Lambda$ scattering}


\author{Yasuhiro~Yamaguchi}
\email[]{yasuhiro.yamaguchi@ge.infn.it}
\affiliation{Yukawa Institute for Theoretical Physics, Kyoto University, Kyoto 606-8502, Japan}
\affiliation{INFN Sezione di Genova, Via Dodecaneso 33, 16146 Genova, Italy}
\author{Tetsuo~Hyodo}
\email[]{hyodo@yukawa.kyoto-u.ac.jp}
\affiliation{Yukawa Institute for Theoretical Physics, Kyoto University, Kyoto 606-8502, Japan}


\date{\today}

\begin{abstract}
We study the quark mass dependence of the H-dibaryon in the strangeness
 $S=-2$ baryon-baryon scattering. A low-energy effective field theory is
 used to describe the coupled-channel scattering, in which the quark
 mass dependence is incorporated so as to reproduce the lattice QCD data
 {by the HAL QCD collaboration} in the SU(3) limit. We point out the existence of the Castillejo-Dalitz-Dyson (CDD) pole in the $\Lambda\Lambda$ scattering amplitude below the threshold in the SU(3) limit, which may cause the Ramsauer-Townsend effect near the $N\Xi$ threshold at the physical point. The H-dibaryon is unbound at the physical point, and a resonance appears just below the $N\Xi$ threshold. As a consequence of the coupled-channel dynamics, the pole associated with the resonance is not continuously connected to the bound state in the SU(3) limit. Through the extrapolation in quark masses, we show that the unitary limit of the $\Lambda\Lambda$ scattering is achieved between the physical point and the SU(3) limit. We discuss the possible realization of the ``H-matter'' in the unphysical quark mass region.
\end{abstract}




\maketitle

\section{Introduction}

The two-baryon system with spin $J=0$, isospin $I=0$ and strangeness $S=-2$ is of particular interest in the strangeness nuclear physics, because of the possible existence of the H-dibaryon. The H-dibaryon was predicted to be stable against the strong decay with the MIT bag model~\cite{Jaffe:1976yi}. A remarkable recent finding by the lattice QCD simulations is that the two-baryon system of these quantum numbers indeed supports a bound state at relatively heavy quark mass region~\cite{Beane:2010hg,Inoue:2010es,Beane:2011iw,Inoue:2011ai,Sasaki:2013zwa,Sasaki:2015ifa,Junnarkar:2015jyf}. On the other hand, at the physical point, 
the existence of the bound H-dibaryon is confronted by a challenge from several experimental data. The observation of the double $\Lambda$ hypernuclei~\cite{Takahashi:2001nm,Ahn:2013poa} excludes the existence of the H-dibaryon with the binding energy larger than $\sim 7$ MeV. The Belle collaboration searched for the H-dibaryon in the $\Upsilon(1S)$ and $\Upsilon(2S)$ decays, finding no clear evidence in the $\Lambda p \pi^{-}$ and $\Lambda\Lambda$ mass spectra~\cite{Kim:2013vym}. The H-dibaryon signal was not found also in the $\Lambda p \pi^{-}$ spectrum from the Pb-Pb collisions at $\sqrt{s_{NN}}=2.76$ TeV performed by the ALICE collaboration at Large Hadron Collider (LHC)~\cite{Adam:2015nca}.
Recently, the STAR collaboration extracted the $\Lambda\Lambda$ correlation function at the Relativistic Heavy-Ion Collider (RHIC)~\cite{Adamczyk:2014vca}. A detailed analysis of the STAR data indicates the attractive scattering length of the $\Lambda\Lambda$ system, as long as the pair purity parameter $\lambda$ is constrained by the measured $\Sigma^{0}/\Lambda$ ratio~\cite{Morita:2014kza,Ohnishi:2015cnu,Ohnishi:2016elb}. The attraction at threshold is consistent with the absence of the bound H-dibaryon below the threshold.

In view of the lattice results and the current status of experimental searches, a plausible scenario is that the H-dibaryon is unbound at the physical point, while it is bound below the $\Lambda\Lambda$ threshold when the quark masses are increased. In other words, there will be a level crossing of the H-dibaryon state and the $\Lambda\Lambda$ state along with the change of the quark masses. This implies the existence of the quark mass region where the $\Lambda\Lambda$ system supports a very shallow bound state with almost zero binding energy, having an infinitely large scattering length. Such a situation is called the unitary limit where various interesting phenomena will take place both in the few-body and many-body systems~\cite{Zwerger,Braaten:2004rn}. In this respect, the H-dibaryon in the $\Lambda\Lambda$ scattering is analogous to the $\sigma$ meson in the $\pi\pi$ scattering where the Efimov effect of three pions is predicted to occur in a certain unphysical quark mass region~\cite{Hyodo:2013zxa}. To examine this possibility for the $\Lambda\Lambda$ system, we need to know how the H-dibaryon in the $\Lambda\Lambda$ scattering behaves with the variation of the quark masses.

The quark mass dependence of the H-dibaryon has been studied by two complementary approaches. One is to evaluate the Nambu-Goldstone (NG) boson loop effect to the flavor singlet bare H-state~\cite{Shanahan:2011su,Shanahan:2013yta}. Another study adopts chiral perturbation theory (ChPT) for baryon-baryon systems with four-point contact interaction and the NG boson exchange contributions~\cite{Haidenbauer:2011ah,Haidenbauer:2011za,Haidenbauer:2015zqb}. The physical picture of the H-dibaryon in the former approach corresponds to the compact six-quark state, while the latter approach deals with the loosely bound baryon-baryon molecular state. In both cases, the H-dibaryon is found to be unbound at the physical point when the lattice QCD data are used to determine the unknown constants.

In general, the dominant non-analytic contribution of the quark mass dependence of hadron masses comes from the chiral loop of the NG boson. In the case of the dibaryon system near the threshold, however, a substantial contribution is expected from the energetically closer two-baryon channels, which are not considered in Refs.~\cite{Shanahan:2011su,Shanahan:2013yta}. The relative importance of the NG boson loop is suppressed when the quark masses are increased, and the correct near-threshold scaling~\cite{Hyodo:2014bda} cannot be reproduced without the coupling to the baryon-baryon channels. ChPT is the standard and systematic tool to study the quark mass dependence of hadrons. However, the available lattice results in the SU(3) limit (the NG boson mass is about 400-800 MeV) may not be in the region where the perturbation theory well converges. In addition, the symmetry argument does not specify the relevant hadronic degrees of freedom other than the NG bosons. For instance, the existence of a bare H-dibaryon field is in principle not excluded.

In this paper, we study the quark mass dependence of the H-dibaryon and
the near-threshold $\Lambda\Lambda$ scattering, with the lattice QCD
data {by the HAL QCD collaboration~\cite{Inoue:2011ai}} being constraints. To this end, we focus on the characteristic length scales in the lattice QCD simulations in the SU(3) limit. In the simulations in Ref.~\cite{Inoue:2011ai}, it is found that the scattering length is larger than $1$ fm~\cite{InouePC}, while the interaction range estimated by the NG-boson exchange $\lambda_{\pi}=1/m_{\pi}$ is at most 0.4 fm.\footnote{Strictly speaking, the pion exchange is absent in the $\Lambda\Lambda$ channel. However, since the pion is the lowest energy excitation in QCD, $\lambda_{\pi}$ can be regarded as the upper limit of the range of the strong interaction.} In such cases, the interaction can be regarded as pointlike and the pionless framework of the effective field theory (EFT) should be valid to describe the near-threshold phenomena~\cite{Kaplan:1996xu,Kaplan:1996nv,vanKolck:1998bw,Bedaque:2002mn,Braaten:2007nq,Epelbaum:2008ga}. We thus construct an EFT to study the H-dibaryon in the two-baryon scattering, as a generalization of the EFT for the nuclear forces. We then introduce the quark mass dependence in the parameters of the EFT, with the lattice QCD result in the SU(3) limit~\cite{Inoue:2011ai} being the guiding principle. This enables us to extrapolate the scattering amplitude with the up and down quark mass $m_{l}$ and the strangeness quark mass $m_{s}$. Preliminary results with only the singlet component interaction can be found in Ref.~\cite{YamaguchiHYP}. Here we present the complete formulation including 8 and 27 components, and the detailed discussion on the behavior of the $\Lambda\Lambda$ scattering amplitude and the structure of H-dibaryon.

This paper is organized as follows. We formulate the EFT for the coupled-channel baryon-baryon scattering in Sec.~\ref{sec:EFT}. The quark mass dependence is discussed in Sec.~\ref{sec:quarkmass}. Combining with the lattice QCD results, we show the results of the quark mass dependence of the baryon-baryon scattering in Sec.~\ref{sec:results}. The last section is devoted to a summary of this work.

\section{Effective field theory}\label{sec:EFT}

In the following, we introduce the low-energy effective field theory for the description of the two-baryon system with $S=-2$, $J=0$, and $I=0$. As long as the small energy region is concerned, the system can be described by the nonrelativistic local quantum field theory with contact interactions~\cite{Kaplan:1996nv,Braaten:2007nq}. In this section, we consider the dynamics of the baryon-baryon scattering for a given set of quark masses. The quark mass dependence of the EFT will be discussed in Sec.~\ref{sec:quarkmass}. We always work in the isospin symmetric limit $m_{u}=m_{d}\equiv m_{l}$, while the SU(3) symmetry may be broken by the strange quark mass, $m_{s}\neq m_{l}$. As shown in Section~\ref{sec:quarkmass}, this causes the SU(3) breaking in baryon masses. 

\subsection{Effective Lagrangian}

We consider the system of the bare H-dibaryon coupled with the two-baryon scattering states. The free part of the Lagrangian density of the nonrelativistic effective field theory is given by
\begin{align}
    \mathcal{L}_{\rm free}
    &= 
    \sum_{a=1}^{4}
    \sum_{\sigma=\uparrow,\downarrow}
    B_{a,\sigma}^{\dag}
    \left(
    i\frac{\partial}{\partial t}
    +
    \frac{\nabla^{2} }{2M_{a}}
    +\delta_{a}
    \right)B_{a,\sigma} \nonumber \\
    &\quad +H^{\dag} 
    \left(i\frac{\partial}{\partial t}
    +\frac{\nabla^{2}}{2M_{H}}+\nu\right) H ,
    \label{eq:Lagrangian}
\end{align}
where $a$ labels the flavor of the baryon ($N$, $\Lambda$, $\Sigma$, $\Xi$) and $\sigma$ denotes the spin of the baryon. We introduce $\delta_{a}=(M_{a}-M_{\Lambda}){\rm c}^{2}$ to account for the mass difference of baryons from $\Lambda$. The bare H-dibaryon state is represented by the field $H$. The parameter $\nu$ represents the energy difference of the bare H-dibaryon and the $\Lambda\Lambda$ threshold.

We use the SU(3) symmetric interaction, and the SU(3) breaking effect is included in $M_{a}$ and $\delta_{a}$ which affect the kinematics of the baryon loop diagrams. The SU(3) symmetric interaction can easily be expressed in the SU(3) basis. We denote the two-baryon system in the total spin $J=0$, strangeness $S=-2$, and isospin $I=0$ channel as
\begin{align}
    D^{(F)}
    &= 
    [BB]_{J=0,S=-2,I=0}^{(F)} ,
\end{align}
where $F$ labels the SU(3) representation. In this sector, only the symmetric representations can contribute, so $F=1,8$ and 27. The interaction Lagrangian is then given by
\begin{align}
    \mathcal{L}_{\rm int}
    &= 
    -g
    [D^{(1)\dag}H + H^{\dag}D^{(1)} ]
    - \lambda^{(1)}
    D^{(1)\dag}D^{(1)} \nonumber \\
    &\quad 
    - \lambda^{(8)}
    D^{(8)\dag}D^{(8)}
    - \lambda^{(27)}
    D^{(27)\dag}D^{(27)}
    \label{eq:Lagint} ,
\end{align}
with the coupling constants $g$ and $\lambda^{(F)}$. Here we assume that the H-dibaryon field is in the flavor singlet representation and there are no bare fields in the $8$ and $27$ sectors. The first term represents the three-point contact interaction of the bare H-dibaryon and two baryons, and the other terms represent the four-point contact interactions of baryons in different flavor representations.

The SU(3) basis can be transformed to the isospin basis as~\cite{Sasaki:2015ifa}, 
\begin{align}
    \begin{pmatrix}
    D^{(1)} \\
    D^{(8)} \\
    D^{(27)}
    \end{pmatrix}
    &= 
    U
    \begin{pmatrix}
    \Lambda\Lambda \\
    N\Xi \\
    \Sigma\Sigma
    \end{pmatrix},\quad
    U=
    \begin{pmatrix}
    -\sqrt{\frac{1}{8}} & \sqrt{\frac{1}{2}} & \sqrt{\frac{3}{8}} \\
    -\sqrt{\frac{1}{5}} & \sqrt{\frac{1}{5}} & -\sqrt{\frac{3}{5}} \\
    \sqrt{\frac{27}{40}} & \sqrt{\frac{3}{10}} & -\sqrt{\frac{1}{40}}
    \end{pmatrix} .
\end{align}
The interaction Lagrangian in the isospin basis can be obtained as
\begin{align}
    \mathcal{L}_{\rm int} 
    &= 
    -g
    \left[
    \begin{pmatrix}
    \Lambda^{\dag}\Lambda^{\dag} & N^{\dag}\Xi^{\dag} & \Sigma^{\dag}\Sigma^{\dag}
    \end{pmatrix}d H + H^{\dag}d^{\dag}
    \begin{pmatrix}
    \Lambda\Lambda \\
    N\Xi \\
    \Sigma\Sigma
    \end{pmatrix} 
    \right] \nonumber \\
    &\quad 
    - 
    \begin{pmatrix}
    \Lambda^{\dag}\Lambda^{\dag} & N^{\dag}\Xi^{\dag} & \Sigma^{\dag}\Sigma^{\dag}
    \end{pmatrix}
    V
    \begin{pmatrix}
    \Lambda\Lambda \\
    N\Xi \\
    \Sigma\Sigma
    \end{pmatrix} ,
\end{align}
where
\begin{align}
    d
    &= \begin{pmatrix}
    -\sqrt{\tfrac{1}{8}} \\
    -\sqrt{\tfrac{1}{2}} \\
    \sqrt{\tfrac{3}{8}} 
    \end{pmatrix}, \quad
    V
    =U^{-1}
    \begin{pmatrix}
    \lambda^{(1)} & & \\
    & \lambda^{(8)} & \\
    & & \lambda^{(27)}
    \end{pmatrix}
    U .
\end{align}

\subsection{Scattering amplitude}

We now consider the baryon-baryon scattering amplitude. In the following, we work in the center-of-mass frame of the two-baryon system and evaluate the on-shell scattering amplitude with the total energy $E$, measured from the $\Lambda\Lambda$ threshold. Because of the phase symmetry in the effective Lagrangian, the two-baryon sector is decoupled from the $N$-baryon sectors with $N\neq 2$, so the two-baryon problem can be solved exactly. It is straightforward to derive Feynman rules and write down the tree-level two-baryon amplitude as
\begin{align}
   \mathcal{A}_{ij}^{\rm tree}(E)
   &=
   -\left(V_{ij}
   +\frac{g^{2}d_i^{\dag}d_j}{E-\nu+i0^+}
   \right)
   \label{eq:Atree} ,
\end{align}
where $i,j$ denotes the channel indices in the isospin basis. The two-baryon scattering amplitude is given by the solution of the Lippmann-Schwinger equation:
\begin{align}
   \mathcal{A}_{ij}(E)
   &= \mathcal{A}_{ij}^{\rm tree}(E)
   -\sum_{k}\mathcal{A}_{ik}^{\rm tree}(E)I_{k}(E)
   \mathcal{A}_{kj}(E) .
\end{align}   
The solution is analytically given by
\begin{align}   
   \mathcal{A}(E)
   &= [(\mathcal{A}^{\rm tree}(E))^{-1}+I(E)]^{-1} ,
\end{align}
where $I_{i}(E)$ is defined by
\begin{align}
    I_{i}(E)
    &=
    \int \frac{d^{3}k}{(2\pi)^{3}}
    \frac{1}{E-\Delta_{i}-\frac{\bm{k}^{2}}{2\mu_{i}}+i0^{+}} ,
\end{align}
with $\mu_{1}=M_{\Lambda}/2$, $\mu_{2} = M_{N}M_{\Xi}/(M_{N}+M_{\Xi})$, $\mu_{3}=M_{\Sigma}/2$, $\Delta_{1}=0$, $\Delta_{2}= \delta_{N}+\delta_{\Xi}$, and $\Delta_{3}= 2\delta_{\Sigma}$. Ultraviolet divergence of the integral is tamed by the sharp cutoff $\Lambda$. The regularized loop function for $E-\Delta_{i}>0$ is given by
\begin{align}
    I_{i}(E)
    &= 
    \frac{\mu_{i}}{\pi^{2}}
    \left(-\Lambda+k_{i}\artanh\frac{\Lambda}{k_{i}}\right) 
    \label{eq:Iiabove} ,\\
    k_{i}
    &= \sqrt{2\mu_{i}(E-\Delta_{i})} .
\end{align}
While the EFT is renormalizable, in this study, we keep the finite cutoff at the momentum scale below which the EFT description is reliable, and determine the coupling constants at this scale. For later convenience, we introduce the loop functions in the first and second Riemann sheet for complex $E$ as
\begin{align}
    I_{i,\text{I}}(E)
    &=\frac{\mu_{i}}{\pi^{2}}
    \Biggl[-\Lambda
    +[2\mu_{i} (E-\Delta_{i})]^{1/2} \nonumber\\
    &\quad \times \artanh\frac{\Lambda}{[2\mu_{i} (E-\Delta_{i})]^{1/2}}
    \Biggr] ,\\
    I_{i,\text{II}}(E)
    &=\frac{\mu_{i}}{\pi^{2}}
    \Biggl[-\Lambda
    +[2\mu_{i} (E-\Delta_{i})]^{1/2}\nonumber \\
    &\quad \times 
    \left(\artanh\frac{\Lambda}{[2\mu_{i} (E-\Delta_{i})]^{1/2}}
    +i\pi\right)
    \Biggr] ,
\end{align}
where the arguments of the complex variables are chosen to be $0\leq \theta < 2\pi$. With three coupled channels, the scattering amplitude is defined on the $2^{3}=8$ sheeted Riemann surface. The Riemann sheet is identified by specifying the choice of I/II loop function for each channel. The most adjacent sheet to the real axis is obtained by choosing I for the closed channels and II for the open channels.

The forward scattering amplitude is given by
\begin{align}
   f_{ii}(E)
   &=  \frac{\mu_{i}}{2\pi}[(\mathcal{A}^{\rm tree}(E))^{-1}+I(E)]^{-1}_{ii} .
\end{align}
The scattering length in the $\Lambda\Lambda$ channel is defined as
\begin{align}
   a^{\Lambda\Lambda}
   &= 
   -f_{11}(E)|_{E\to 0}  .
   \label{eq:LLslength}
\end{align}
In this convention, the negative (positive) scattering length stands for the attraction (repulsion) at the threshold. 
{The $N\Xi$ scattering length is given by
\begin{align}
   a^{N\Xi}
   &= 
   -f_{22}(E)|_{E\to \delta_{N}+\delta_{\Xi}}  .
   \label{eq:NXislength}
\end{align}
}

\subsection{SU(3) limit}

In the SU(3) limit $m_{l}=m_{s}$, there is no mass difference in the flavor multiplet, and we denote the baryon mass $M$ and the reduced mass $\mu=M/2$. Because the interaction Lagrangian~\eqref{eq:Lagint} is SU(3) symmetric, the baryon-baryon scattering reduces to the independent single-channel problems. The scattering amplitude in the flavor singlet channel is given by
\begin{align}
   f^{(1)}(E)
   &= 
   \Biggl[-\frac{2\pi}{\mu}\left(\lambda^{(1)}
   +\frac{g^{2}}{E-\nu+i0^+}
   \right) ^{-1} \nonumber \\
   &\quad -\frac{2}{\pi}
   \left(\Lambda-k\artanh\frac{\Lambda}{k}\right)
   \Biggr]^{-1} ,
\end{align}
where $k=\sqrt{2\mu E}$. The amplitudes in the octet and 27-plet channels are given by
\begin{align}
   f^{(8),(27)}(E)
   &= 
   \left[-\frac{2\pi}{\mu\lambda^{(8),(27)}}
   -\frac{2}{\pi}
   \left(\Lambda-k\artanh\frac{\Lambda}{k}\right)
   \right]^{-1} .
\end{align}
The scattering length $a^{(F)}=-f^{(F)}(E=0)$ is given by
\begin{align}
   a^{(1)}
   & =\frac{M}{4\pi}
   \left[
   \left(\lambda^{(1)}
   -\frac{g^{2}}{\nu}
   \right)^{-1}+\frac{M\Lambda}{2\pi^{2}}
   \right]^{-1} 
   \label{eq:a0} , \\
   a^{(8),(27)}
   &= \frac{M}{4\pi}
   \left[
   \frac{1}{\lambda^{(8),(27)}}
   +\frac{M\Lambda}{2\pi^{2}}
   \right]^{-1}  .
\end{align}

\section{Quark mass dependence}\label{sec:quarkmass}

Our aim is to consider the quark mass dependence of the H-dibaryon and the two-baryon scattering amplitude. In the previous section, we introduce the EFT to describe the near-threshold phenomena accurately, but the framework is not based on a systematic expansion with respect to the quark mass. The quark mass dependence should therefore be included in the parameters of the scattering amplitude. 

To begin with, we define the ``quark masses'' $m_{l}$ and $m_{s}$ from the meson masses as (see also Ref.~\cite{Shanahan:2011su})
\begin{align}
   B_{0}m_{l}
   &= 
   \frac{m_{\pi}^{2}}{2} ,\quad
   B_{0}m_{s}
   = 
   m_{K}^{2}-\frac{m_{\pi}^{2}}{2} .
   \label{eq:quarkmass}
\end{align}
Choosing the constant $B_{0}=-\langle \bar{q}q\rangle/(3F_{0}^{2})$ with the quark condensate $\langle \bar{q}q\rangle$ and the pion decay constant $F_{0}$ in the chiral limit, we obtain the Gell-Mann--Oakes--Renner relation~\cite{GellMann:1968rz}. Up to the linear order in quark masses, these are the rigorous relations in QCD to relate the NG boson masses with the quark masses. Of course, it is not always guaranteed that the leading order result works well in the unphysical quark mass region, but it turns out that Eq.~\eqref{eq:quarkmass} is sufficient for the accuracy required in the present study. Higher order corrections in quark masses could be systematically included in ChPT~\cite{Scherer:2012xha}.

We consider the H-dibaryon and the baryon-baryon interaction in $m_{l}$-$m_{s}$ plane. In the following, we introduce the quark mass dependence in the hadron masses and coupling constants. We here consider the minimal dependence up to linear order in $m_{l}$ and $m_{s}$, for the consistency with Eq.~\eqref{eq:quarkmass}.

\subsection{Hadron masses}\label{subsec:massdep}

The baryon masses are expressed in the leading order ChPT as~\cite{Shanahan:2011su}
\begin{align}
   M_{N}(m_{l},m_{s})
   &= 
   M_{0}
   -(2\alpha+2\beta+4\sigma)B_{0}m_{l}
   -2\sigma B_{0}m_{s} , \label{eq:MN} \\
   M_{\Lambda}(m_{l},m_{s})
   &= 
   M_{0}
   -(\alpha+2\beta+4\sigma)B_{0}m_{l}
   \nonumber \\
   &\quad -(\alpha+2\sigma) B_{0}m_{s} ,
   \label{eq:MLambda}\\
   M_{\Sigma}(m_{l},m_{s})
   &= 
   M_{0}
   -\left(\frac{5}{3}\alpha+\frac{2}{3}\beta+4\sigma\right)B_{0}m_{l} \nonumber \\
   &\quad -\left(\frac{1}{3}\alpha+\frac{4}{3}\beta+2\sigma\right) B_{0}m_{s}, \label{eq:MSigma}\\
   M_{\Xi}(m_{l},m_{s})
   &= 
   M_{0}
   -\left(\frac{1}{3}\alpha+\frac{4}{3}\beta+4\sigma\right)B_{0}m_{l}
   \nonumber \\
   &\quad -\left(\frac{5}{3}\alpha+\frac{2}{3}\beta+2\sigma\right) B_{0}m_{s} \label{eq:MXi}, 
\end{align}
with parameters $M_{0}$, $\alpha$, $\beta$, and $\sigma$. 
We note that the combination $M_{0}-2\sigma B_{0}(2m_{l}+m_{s})$ is common for all baryons, and three mass differences are expressed by two parameters, $\alpha$ and $\beta$. This leads to the constraint on the mass differences, known as the Gell-Mann--Okubo formula~\cite{Gell-Mann:1962xb,Okubo:1962jc}
\begin{align}
   \frac{M_{N}+M_{\Xi}}{2}
   =\frac{3M_{\Lambda}+M_{\Sigma}}{4} ,
\end{align}
which is known to be satisfied by the physical ground state baryons at 1\% accuracy. In the SU(3) limit ($m_{l}=m_{s}$), all the baryons have the same mass
\begin{align}
   M_{B}(m_{l})
   &= 
   M_{0}
   -(2\alpha+2\beta+6\sigma)B_{0}m_{l} .
\end{align}

In Eq.~\eqref{eq:Lagrangian}, the parameter $\nu$ represents the energy difference of the bare H-dibaryon and the $\Lambda\Lambda$ threshold. Because the H-dibaryon is introduced as flavor singlet, the bare mass $M_{H}$ should be proportional to the combination $2m_{l}+m_{s}$. Thus, we parametrize the quark mass dependence of $\nu$ by introducing two parameters $M_{H,0}$ and $\sigma_{H}$ as
\begin{align}
   \nu(m_{l},m_{s})/{\rm c}^{2}
   &= 
   M_{H,0}-\sigma_{H}B_{0} \left(2m_{l}+m_{s}\right) \nonumber \\
   &\quad -2M_{\Lambda}(m_{l},m_{s}) ,
\end{align}
where $M_{\Lambda}(m_{l},m_{s})$ is given in Eq.~\eqref{eq:MLambda}.

\subsection{Coupling constants}

There are coupling constants $\lambda^{(F)}$ (four-point vertices, $F=1,8,27$) and $g$ (three-point vertex in the singlet channel) in the effective Lagrangian. In general, we can introduce the quark mass dependence in all coupling constants. We here introduce the linear quark mass dependence in $\lambda^{(F)}$ as
\begin{align}
   \lambda^{(F)}(m_{l},m_{s})
   &= 
   \lambda_{0}^{(F)}+\lambda_{1}^{(F)}B_{0}\left(2m_{l}+m_{s}\right) .
\end{align}
The quark mass dependence is SU(3) symmetric, because the SU(3) breaking term also induces the off-diagonal coupling in SU(3) basis. The three-point vertex is kept as constant:
\begin{align}
   g(m_{l},m_{s})
   &= 
   g,
\end{align}
because the quark mass dependence is induced in the bare-H propagator through $\nu$ in Eq.~\eqref{eq:Atree}.

\section{Numerical results}\label{sec:results}

To determine the quark mass dependence in the EFT, we utilize the HAL QCD results in SU(3) symmetric limit~\cite{Inoue:2011ai} with three lightest quark masses, which we denote HAL-1, HAL-2, and HAL-3. To estimate the systematic uncertainty, we examine two cases in the flavor singlet channel: the ``contact'' model where the coupling to the bare field is switched off ($g=0$), and the ``bare H'' model which includes both the contact interaction and the bare H term. All the coupling constants are given at the fixed cutoff $\Lambda=300$ MeV/c.

\subsection{Baryon masses}\label{sec:baryonfit}

\begin{figure}[tbp]
    \centering
    \includegraphics[width=8cm]{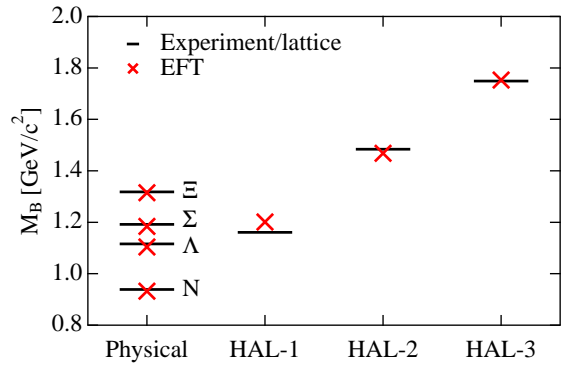}
    \caption{\label{fig:baryonmass}
    (Color online) Baryon masses with Eq.~\eqref{eq:average}. Horizontal bars show the central values of the experimental data and the lattice results from Ref.~\cite{Inoue:2011ai}.}
\end{figure}

To connect the SU(3) limit with the physical point, we first determine $\alpha$ and $\beta$ by mass differences at physical point, and then determine $M_{0}$ and $\sigma$ combining with the lattice results of the baryon masses in the SU(3) limit. The best fit values are obtained as
\begin{align}
   M_{0}
   &=0.948 \text{ [GeV\;c$^{-2}$]},\quad
   \alpha
   =-0.754 \text{ [GeV$^{-1}$c$^{2}$]}, \nonumber \\
   \beta
   &=-0.644 \text{ [GeV$^{-1}$c$^{2}$]},\quad
   \sigma
   =0.0826 \text{ [GeV$^{-1}$c$^{2}$]} .
   \label{eq:average}
\end{align}
The baryon masses with these parameters are compared with the experimental data and the lattice results in Fig.~\ref{fig:baryonmass}. Physically, we expect that the coefficients in front of $m_{l}$ and $m_{s}$ should be positive, because the baryon mass should increase along with the quark mass. This is guaranteed when $\alpha$, $\beta$, and $\sigma$ are all negative. Although we obtain the solution with $\sigma>0$, it is confirmed that all the baryon masses increase with the quark masses, except for the $m_{s}$ dependence of the nucleon mass. 

\subsection{Coupling constants in 1 channel}

\begin{figure}[tbp]
    \centering
    \includegraphics[width=8cm]{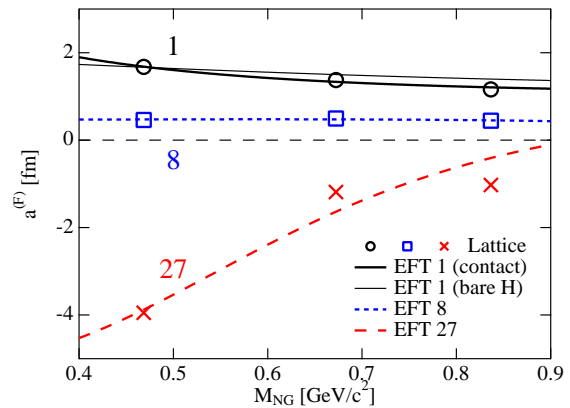}
    \caption{\label{fig:slength}
    (Color online) Scattering lengths of the baryon-baryon scattering in the SU(3) limit by the lattice QCD simulation~\cite{Inoue:2011ai,InouePC} and by the EFT. Circles (solid lines), squares (dotted line), and crosses (dashed line) denote the lattice (EFT) results in the flavor 1 channel, 8 channel, and 27 channel, respectively.
    In the singlet channel, thick (thin) line represents the result in the contact (bare H) model. }
\end{figure}%

\begin{table}[tbp] 
	\begin{center}
		\begin{ruledtabular}
\begin{tabular}{cccc}
Data & Lattice~\cite{Inoue:2011ai}  & Contact model & Bare H model \\
\hline
HAL-3 & 38 & 89 & 38 \\ 
HAL-2 & 34 & 57 & 33 \\
HAL-1 & 26 & 26 & 27 \\ 
\end{tabular}
\caption{Binding energies in the flavor 1 channel in units of MeV\;c$^{-2}$. Lattice results are taken from Ref.~\cite{Inoue:2011ai}. }
\label{tbl:bindingenergy}
        \end{ruledtabular}
	\end{center}
\end{table}%

We first present the parameters in the flavor singlet channel. In the contact model where $g=0$, we determine the coupling constants $\lambda_{0}^{(1)}$ and $\lambda_{1}^{(1)}$ by the scattering lengths in the SU(3) limit obtained by the HAL QCD collaboration~\cite{Inoue:2011ai,InouePC}. The best fit values are obtained as
\begin{align}
   \lambda_{0}^{(1)}
   &=-88.5 \text{ [GeV$^{-2}$c$^{3}$]},\quad
   \lambda_{1}^{(1)}
   =-163 \text{ [GeV$^{-4}$c$^{7}$]} .
   \label{eq:1couplingnoH} 
\end{align}
The resulting scattering lengths (thick solid line) are compared with the lattice results (circles) in Fig.~\ref{fig:slength}. The lattice results of the scattering lengths are  well reproduced. We note that both  $\lambda_{0}^{(1)}$ and $\lambda_{1}^{(1)}$ are negative in Eq.~\eqref{eq:1couplingnoH}. Namely, the singlet interaction is attractive, while the scattering lengths are positive. In fact, we find a bound state as summarized in Table~\ref{tbl:bindingenergy}. While the existence of the bound state is qualitatively consistent with lattice QCD, the binding energies deviate from the lattice results, in particular in the heavier quark mass case. Because the contact interaction model is reliable at the threshold energy, it is reasonable to determine the coupling constants by the scattering lengths.

In the bare H model, the applicable energy region is slightly increased by the presence of the pole term. Because there are three additional parameters, $g$, $M_{H,0}$, and $\sigma_{H}$, we use the binding energies obtained by the HAL QCD collaboration~\cite{Inoue:2011ai} to determine the parameters. We set $\lambda_{1}^{(1)}=0$ so that the quark mass dependence is governed by the bare H term. We then obtain
\begin{align}
   \lambda_{0}^{(1)}
   &=-12.8 \text{ [GeV$^{-2}$c$^{3}$]},\quad
   g^{2}
   =2350 \text{ [GeV$^{-1}$c$^{2}$]} , \nonumber \\
   M_{H,0}
   &=19.8 \text{ [GeV\;c$^{-2}$]} ,\quad 
   \sigma_{H}
   =-1.53 \text{ [GeV$^{-1}$c$^{2}$]} .
   \nonumber
\end{align}
The results of the scattering length and the binding energy are shown in Fig.~\ref{fig:slength} and Table~\ref{tbl:bindingenergy}. In the bare H model, both quantities are well reproduced. The large value of $M_{H,0}$ is worth mentioning; the constraint from the lattice QCD data excludes the existence of the bare H field near the two-baryon threshold. Because it is an order of magnitude larger than $2M_{\Lambda}$, in the low-energy region, the pole term of the bare H state only produces a smooth energy dependence. Thus, the pole term in the bare H model can be regarded as a higher-order correction to the four-point contact term.

\subsection{Coupling constants in 8 and 27 channels}

We determine the coupling constants $\lambda_{0}^{(F)}$ and $\lambda_{1}^{(F)}$ for $F=8$ and 27 by the scattering lengths in the SU(3) limit obtained by the HAL QCD collaboration~\cite{Inoue:2011ai,InouePC}. The best fit values are given by
\begin{align}
   \lambda_{0}^{(8)}
   &=54.2 \text{ [GeV$^{-2}$c$^{3}$]},\quad
   \lambda_{1}^{(8)}
   =-23.7 \text{ [GeV$^{-4}$c$^{7}$]} ,
   \nonumber \\
   \lambda_{0}^{(27)}
   &=-58.2 \text{ [GeV$^{-2}$c$^{3}$]},\quad
   \lambda_{1}^{(27)}
   =45.3 \text{ [GeV$^{-4}$c$^{7}$]} .
   \nonumber
\end{align}
In the flavor 8 channel, the scattering lengths are relatively well
reproduced, as shown in Fig.~\ref{fig:slength}. We check that these parameters provide repulsive (attractive) interaction in the flavor 8 (27) channel, $\lambda_{0}^{(8)}+\lambda_{1}^{(8)}B_{0}(2m_{l}+m_{s})>0$ ($\lambda_{0}^{(27)}+\lambda_{1}^{(27)}B_{0}(2m_{l}+m_{s})<0$) in the quark mass region of the lattice data. This is consistent with the absence of the bound state in these channels. 

\subsection{$\Lambda\Lambda$ scattering in the SU(3) limit}

\begin{figure}[tbp]
    \centering
    \includegraphics[width=8cm]{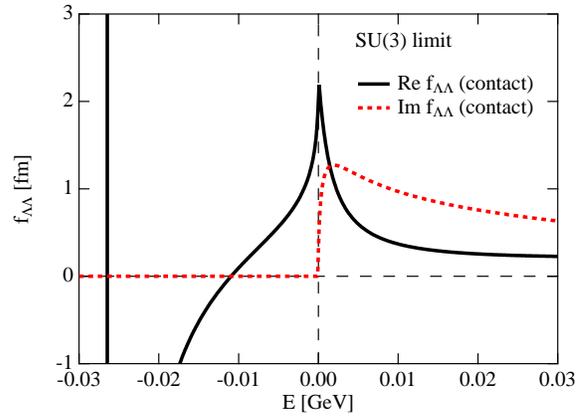}
    \caption{\label{fig:LLampSU3}
    (Color online) $\Lambda\Lambda$ scattering amplitude in the contact model in the SU(3) limit. 
    }
\end{figure}

Using the parameters determined above, we study the quark mass dependence of the $\Lambda\Lambda$ scattering. We first discuss the $\Lambda\Lambda$ scattering amplitude in the SU(3) symmetric limit at HAL-1. In the SU(3) limit, the $\Lambda\Lambda$ scattering amplitude is given by the linear combination of the amplitudes in the SU(3) basis as
\begin{align}
   f^{\Lambda\Lambda}(E)
   &=  \frac{1}{8}f^{(1)}(E)
   +\frac{1}{5}f^{(8)}(E)
   +\frac{27}{40}f^{(27)}(E) .
   \label{eq:LLampSU3}
\end{align}
The scattering amplitude of the contact model is shown in Fig.~\ref{fig:LLampSU3}. In the SU(3) limit, the bare H model gives the almost identical amplitude in the energy region of Fig.~\ref{fig:LLampSU3}, because the scattering length are fitted to the same data in both cases. We note that the $\Lambda\Lambda$ scattering length in Eq.~\eqref{eq:LLslength} is attractive,
\begin{align}
   a^{\Lambda\Lambda}
   &= 
   \begin{cases} -2.31 \text{ [fm]} & \text{(contact, SU(3) limit)} \\
   -2.32 \text{ [fm]} & \text{(bare H, SU(3) limit)}
   \end{cases} ,
   \label{eq:slengthSU3}
\end{align}
while there is a bound state below the threshold at $E=-26$ ($-27$) MeV in the contact (bare H) model. The result is qualitatively consistent with the calculation by the HAL QCD collaboration with the lattice QCD potential (see Fig.~8 in Ref.~\cite{Inoue:2011ai}). 

\begin{figure}[tbp]
    \centering
    \includegraphics[width=8cm]{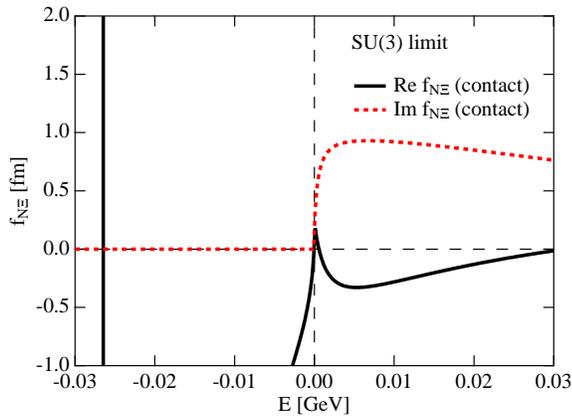}
    \caption{\label{fig:NXiampSU3}
    (Color online) $N\Xi$ scattering amplitude in the contact model in the SU(3) limit. 
    }
\end{figure}%

At first glance, the attractive scattering length in Eq.~\eqref{eq:slengthSU3} is somehow counterintuitive, because the scattering length would be repulsive if there is a shallow bound state, according to the low-energy universality~\cite{Braaten:2004rn}. Let us consider the origin of this structure. From Eq.~\eqref{eq:LLampSU3}, the amplitude $f^{\Lambda\Lambda}(E)$ should have a bound state at the energy of the bound state in the singlet amplitude $f^{(1)}(E)$. On the other hand, the linear combination in Eq.~\eqref{eq:LLampSU3} puts the largest weight in $f^{(27)}(E)$, which has a large attractive scattering length as shown in Fig.~\ref{fig:slength}. This leads to the negative $\Lambda\Lambda$ scattering length. This means that $f^{\Lambda\Lambda}(E=0)>0$ and $\lim_{\epsilon\to +0}f^{\Lambda\Lambda}(-B+\epsilon)\to -\infty$ where $B>0$ is the binding energy.

As a consequence, between the threshold and the bound state pole, there is an energy at which the amplitude vanishes. The zero of the scattering amplitude is called the Castillejo-Dalitz-Dyson (CDD) pole~\cite{Castillejo:1956ed}. Because the effective range expansion is the expansion of the inverse amplitude $[f(E)]^{-1}$, its convergence radius cannot go beyond the closest CDD pole from the threshold. In this way, the attractive scattering length at threshold can coexist with the shallow bound state below the threshold, thanks to the CDD pole between them. We emphasize that this structure is caused by the mixing of the amplitude with a bound state (flavor 1 channel) and that with an attractive scattering length (flavor 27 channel). In other words, the coupled-channel effect plays an important role for the appearance of the CDD pole near the threshold.

{
We also calculate the $N\Xi$ scattering amplitude in the SU(3) limit at HAL-1,
which is given by
\begin{align}
 f^{N\Xi}(E)=\frac{1}{2}f^{(1)}(E) + \frac{1}{5}f^{(8)}(E) +
 \frac{3}{10}f^{(27)}(E) ,
\end{align}
in the SU(3) basis.
The results of the contact model are shown in Fig.~\ref{fig:NXiampSU3}.
The existence of a bound state is seen in
Fig.~\ref{fig:NXiampSU3}, which originates in the bound state in the
amplitude $f^{(1)}(E)$ in the same way as the $\Lambda\Lambda$
scattering amplitude.
The real part of the $N\Xi$ amplitude is a small positive value at $E=0$.
Therefore, a small attractive scattering length is obtained as
\begin{align}
   a^{N\Xi}
   &= 
   \begin{cases} -0.23 \text{ [fm]} & \text{(contact, SU(3) limit)} \\
   -0.24 \text{ [fm]} & \text{(bare H, SU(3) limit)}
   \end{cases} .
   \label{eq:slengthNXiSU3}
\end{align}
As seen in the $\Lambda\Lambda$ scattering, 
the large contribution from the 27 channel makes the $N\Xi$ scattering
length attractive. 
The small attractive scattering length of the $N\Xi$ channel is consistent
with HAL QCD results~\cite{InouePC}.
}

\subsection{$\Lambda\Lambda$ scattering at physical point}

\begin{figure}[tbp]
    \centering
    \includegraphics[width=8cm]{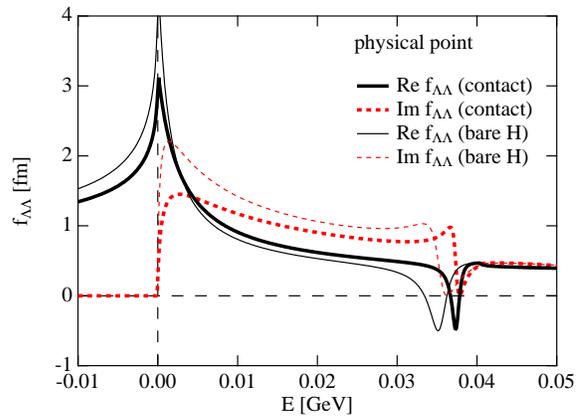}
    \caption{\label{fig:LLampphys}
    (Color online) $\Lambda\Lambda$ scattering amplitude at the physical point.
    Thick (thin) lines represent the results in the contact (bare H) model. 
    }
\end{figure}%

\begin{figure}[tbp]
    \centering
    \includegraphics[width=8cm]{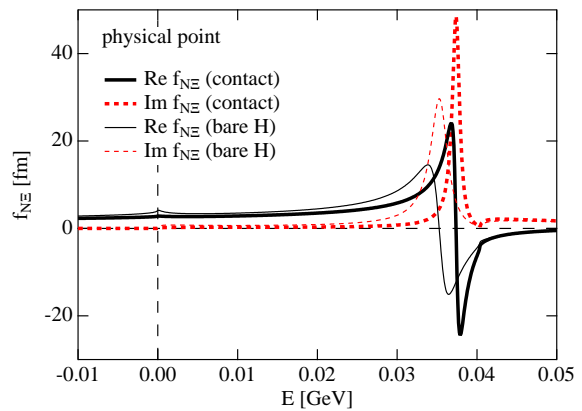}
    \caption{\label{fig:NXiampphys}
    (Color online) $N\Xi$ scattering amplitude at the physical point.
 Same convention as Fig.~\ref{fig:LLampphys}.
    }
\end{figure}%

Next, we calculate the $\Lambda\Lambda$ scattering amplitude at the physical point. We use the physical values of the pion and kaon masses $m_{\pi,K}^{\text{phys}}$ to calculate the baryon masses and the coupling constants. The results are shown in Fig.~\ref{fig:LLampphys}. While the quantitative deviation of the two models is now evident, the qualitative behavior of the amplitude is similar with each other. In both cases, we have checked that no bound state is found below the threshold, and the scattering length is attractive:\footnote{The result presented in Ref.~\cite{YamaguchiHYP} is obtained only with the singlet component. By adding the 8 and 27 components, we obtain the result of the bare H model in Eq.~\eqref{eq:slengthphys}.}
\begin{align}
   a^{\Lambda\Lambda}
   &=  
   \begin{cases}
   -3.22 \text{ [fm]} & \text{(contact, physical point)} \\
   -4.71 \text{ [fm]} & \text{(bare H, physical point)}
   \end{cases} .
   \label{eq:slengthphys}
\end{align}
This is consistent with the absence of the bound H dibaryon at the
physical point, along the same line with the experimental results,
and previous studies of the chiral extrapolation of the lattice QCD data in
Refs.~\cite{Inoue:2011ai,Shanahan:2011su,Shanahan:2013yta,Haidenbauer:2011ah,Haidenbauer:2011za}.
A large magnitude of the scattering length supports the validity of the
extrapolation using the pionless EFT framework.
While
the magnitude of the scattering length is larger than expected (for
instance, $a^{\Lambda\Lambda}=-(0.6$-0.7) fm in next-to-leading order
ChPT~\cite{Haidenbauer:2015zqb}), the attractive nature of the
scattering length is qualitatively reproduced.

We also call attention to the structure near the $N\Xi$
threshold.\footnote{The physical $N\Xi$ threshold is at $E=25$ MeV,
while it appears at $E=40$ MeV in the present calculation. This
difference is caused by the simple linear extrapolation formulae of
baryon masses in Sec.~\ref{subsec:massdep}.} The imaginary part of the
amplitude shows a small peak structure. Associated with this structure,
there is a resonance pole in the complex energy plane with the Riemann
sheet unphysical for the $\Lambda\Lambda$ channel and physical for the
other channels (hereafter called II-I-I sheet). The pole position is
found at $E=37-0.6i$ MeV in the contact model and $E=35-1.3i$ MeV in the
bare H model. It turns out that the residue of this pole in the
$\Lambda\Lambda$ channel is so small that the peak structure on the real
axis is not very prominent, even though the pole locates in the vicinity
of the real axis. In contrast, the residue in the $N\Xi$ channel is much
larger than that in the $\Lambda\Lambda$ channel 
{as seen in the $N\Xi$ scattering amplitude shown in
Fig.~\ref{fig:NXiampphys}}.
{The $N\Xi$ scattering amplitude has a large peak structure, and
gives a scattering length,
\begin{align}
 a^{N\Xi}
   &=  
   \begin{cases}
   3.86 - 0.30i \text{ [fm]} & \text{(contact, physical point)} \\
   3.08 - 0.30i \text{ [fm]} & \text{(bare H, physical point)}
   \end{cases} ,
   \label{eq:slengthNXiphys}
\end{align}
with a complex number because of the existence of the open
$\Lambda\Lambda$ threshold.
 } 
This indicates the
interpretation of this resonance as a $N\Xi$ quasibound state. 
{
We compare the present results with previous works of the chiral extrapolation. In Refs.~\cite{Haidenbauer:2011ah,Haidenbauer:2011za}, the bound state in the unphysical quark mass region is shown to become unbound at the physical point. However, the fate of the bound state at the physical point depends on the choice of the lattice constraints; the HAL QCD results indicate that the resonance disappears, while the NPLQCD result shows the $N\Xi$ quasibound state. In this way, the chiral extrapolation of the lattice results still suffers from systematic uncertainties. We note that, in contrast to the present study, the $8$ and $27$ interactions are constrained by the experimental data of baryon scattering in Refs.~\cite{Haidenbauer:2011ah,Haidenbauer:2011za}. The inclusion of the baryon scattering data in the present framework will be an interesting future direction.
}

In 
{the $\Lambda\Lambda$ scattering amplitude in Fig.~\ref{fig:LLampphys},}
between the resonance peak and the $N\Xi$ threshold, there is a point where both the real and imaginary parts of the amplitude vanish. This occurs when the phase shift passes through $\delta= \pi$, because the s-matrix at this point cannot be distinguished from the noninteracting scattering with $\delta=0$. If the $s$-wave phase shift crosses $\delta= \pi$ at sufficiently low energy where the higher partial waves are negligible, the \textit{total} cross section should almost vanish, like the Ramsauer-Townsend effect~\cite{Taylor}. It is an interesting possibility that the $\Lambda\Lambda$ scattering undergoes the Ramsauer-Townsend effect below the $N\Xi$ threshold.
As we discussed in the previous subsection, the vanishing of the amplitude is also attributed to the CDD pole. In contrast to the SU(3) limit case, here the CDD pole appears in the physical scattering region $E>0$. The Ramsauer-Townsend effect is also discussed in the $\pi\pi$ scattering near the $f_{0}(980)$ resonance in Ref.~\cite{Harada:1995dc}. A similar behavior of the $\pi\Sigma\to\pi\Sigma$ amplitude near the $\bar{K}N$ threshold is discussed in connection with the structure of the $\Lambda(1405)$ resonance~\cite{Kamiya:2016oao} (see also Refs.~\cite{Hyodo:2007jq,Hyodo:2011ur,Ikeda:2012au} for the $\pi\Sigma\to\pi\Sigma$ amplitude).

\subsection{Interpolation of physical point and SU(3) limit}\label{subsec:interpolation}

\begin{figure}[tbp]
    \centering
    \includegraphics[width=8cm]{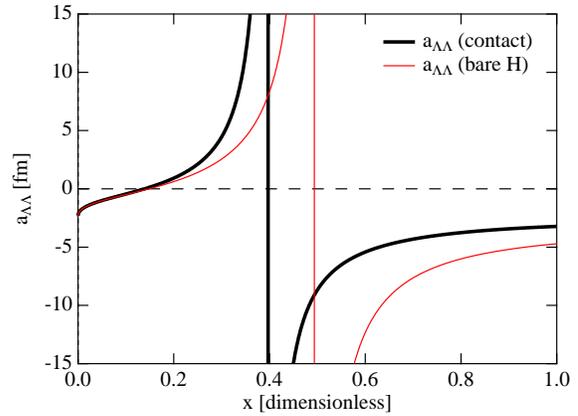}
    \caption{\label{fig:slengthextra}
    (Color online) $\Lambda\Lambda$ scattering length as a function of the interpolation parameter $x$ in Eq.~\eqref{eq:extra1}. $x=0$ ($x=1$) corresponds to the HAL-1 data in the SU(3) limit (physical point).
     Thick (thin) lines represent the results in the contact (bare H) model. 
     }
\end{figure}%


Combining with the bound H-dibaryon in the SU(3) limit, the absence of the bound state at the physical point result indicates the existence of the unitary limit of the $\Lambda\Lambda$ scattering between the physical point and the SU(3) limit. To illustrate this, let us make an interpolation of the physical point and the HAL-1 point in the SU(3) limit by
\begin{align}
   m_{\pi,K}(x)
   &= xm_{\pi,K}^{\text{phys}}+(1-x)m_{\pi,K}^{\text{HAL-1}} ,
   \label{eq:extra1}
\end{align}
where $x=0$ ($x=1$) corresponds to the SU(3) limit (physical point). The
$\Lambda\Lambda$ scattering length as a function of $x$ is shown in
Fig.~\ref{fig:slengthextra}. We see that the unitary limit is indeed
realized at $x\sim 0.4$ in the contact model and $x\sim 0.5$ in the bare
H model where the scattering length diverges. 

\begin{figure}[tbp]
    \centering
    \includegraphics[width=8cm]{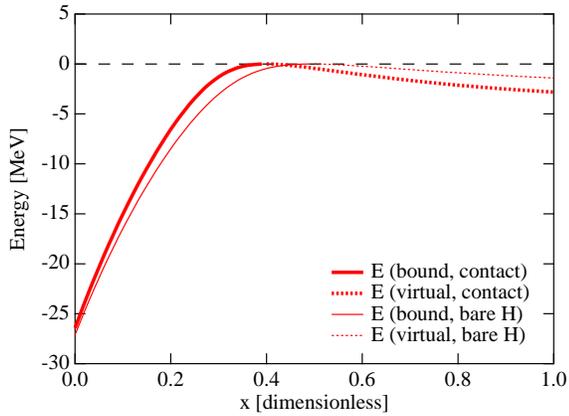}
    \caption{\label{fig:polexBV}
    (Color online) Behaviors of the eigenenergies as functions of the interpolation parameter $x$ in Eq.~\eqref{eq:extra1}. The bound (virtual) state is shown by the solid (dashed) line. Thick (thin) line represents the result in the contact (bare H) model. 
    }
\end{figure}%

\begin{figure}[tbp]
    \centering
    \includegraphics[width=8cm]{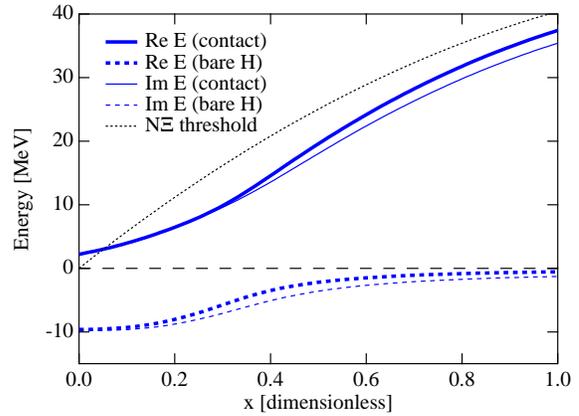}
    \caption{\label{fig:polexR}
    (Color online) Behaviors of the eigenenergies as functions of the interpolation parameter $x$ in Eq.~\eqref{eq:extra1}. The real (imaginary) part is shown by the solid (dashed) line. Thick (thin) line represents the result in the contact (bare H) model. The dotted line stands for the energy of the $N\Xi$ threshold.
    }
\end{figure}

Next question is whether the resonance near the $N\Xi$ threshold at the physical point originates in the bound state in the SU(3) limit. For this purpose, we now study the trajectory of the pole in the analytically continued scattering amplitude, which represents the eigenstate of the Hamiltonian. As already mentioned, we have a bound state at $x=0$ in the SU(3) limit. The corresponding pole is on the physical Riemann sheet for all the channels (I-I-I sheet). The behavior of the eigenenergy as a function of $x$ is shown in Fig.~\ref{fig:polexBV}. As we increase $x$, the binding energy reduces, and eventually vanishes at the unitary limit. The pole then turns into a virtual state on the II-I-I sheet where the $\Lambda\Lambda$ channel is unphysical, in accordance with the general threshold scaling law~\cite{Hyodo:2014bda}. As we further increase $x$, the pole stays below the threshold up to $x=1$. Next, we follow the resonance pole in the II-I-I sheet at the physical point by decreasing $x$. The real and imaginary parts of the eigenenergy are shown by the solid and dashed lines in Fig.~\ref{fig:polexR}. With the decrease of $x$, the real part of the eigenenergy decreases and the magnitude of the imaginary part increases. We note that the energy of the $N\Xi$ threshold measured from the $\Lambda\Lambda$ threshold also decreases, as indicated by the dotted line in Fig.~\ref{fig:polexR}. Around $x\sim 0.17$, the $N\Xi$ threshold becomes lower than the real part of the eigenenergy. At $x=0$, the pole remains above the threshold in the II-I-I Riemann sheet. In this way, we find that the bound state in the SU(3) limit is \textit{not} continuously connected to the resonance found at the physical point. 

\begin{figure}[tbp]
    \centering
    \includegraphics[width=8cm]{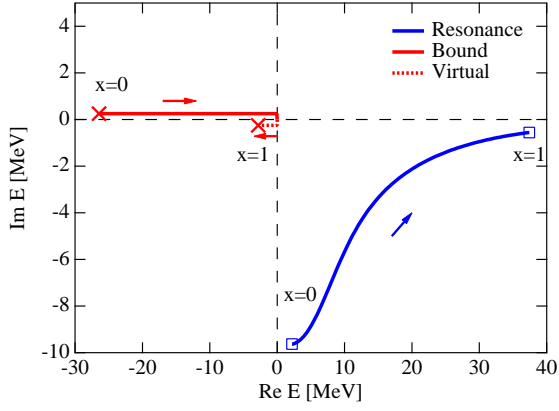}
    \caption{\label{fig:pole}
    (Color online) Trajectories of the poles in the $\Lambda\Lambda$ scattering in the contact model with a variation of the interpolation parameter $x$ in Eq.~\eqref{eq:extra1}. The arrows indicate the direction of the pole movement with the increase of the parameter $x$.
    The bound state (virtual and resonance state) pole is on the I-I-I (II-I-I) sheet. Bound and virtual poles are slightly shifted from the real axis for the purpose of illustration.}
\end{figure}

The pole trajectories in the complex energy plane are illustrated in Fig.~\ref{fig:pole}. At $x=0$, we have a bound state pole below the threshold in the I-I-I sheet and a pole above the threshold in the II-I-I sheet. As we increase the parameter $x$, the former evolves to a virtual state while the latter becomes a resonance near the $N\Xi$ threshold. Relatively large $\Lambda\Lambda$ scattering lengths in Eq.~\eqref{eq:slengthphys} can be understood by the existence of a virtual pole below the threshold. It should be noted that the pole in the II-I-I sheet with $x=0$ is not identified as a resonance state, because the most adjacent Riemann sheet to the real axis in the SU(3) limit is the II-II-II sheet where no resonance pole is found. Rather, we may identify the pole in the II-I-I Riemann sheet as a shadow pole of the bound state in the SU(3) limit~\cite{Eden:1964zz}. At $x=1$, the pole appears in the II-I-I sheet together with its shadow pole, as observed in the $N_{c}$ scaling analysis of the $\Lambda(1405)$~\cite{Hyodo:2007np,Roca:2008kr}. Similar pole trajectories with Fig.~\ref{fig:pole} are expected in the extrapolation performed in Ref.~\cite{Inoue:2011ai} where the hadron masses are extrapolated to the physical values with the lattice potential in the SU(3) limit. Unfortunately, the complex scaling method used in Ref.~\cite{Inoue:2011ai} cannot find the virtual state pole and the pole below the $N\Xi$ threshold in the unphysical Riemann sheet.

\begin{figure}[tbp]
    \centering
    \includegraphics[width=8cm]{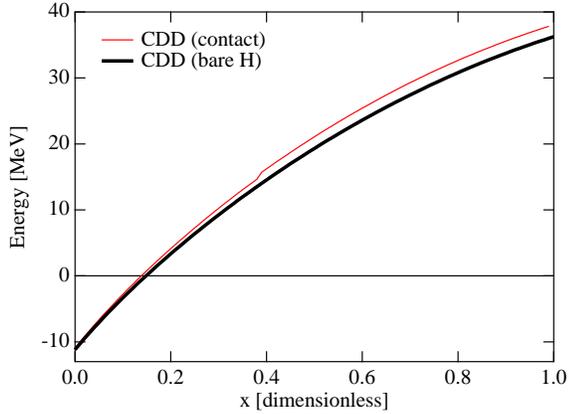}
    \caption{\label{fig:CDD}
    (Color online) Behaviors of the energies of the CDD pole as functions of the interpolation parameter $x$ in Eq.~\eqref{eq:extra1}. Thick (thin) line represents the result in the contact (bare H) model. }
\end{figure}

The behavior of the CDD pole (zero of the $\Lambda\Lambda$ amplitude) is also worth investigating. In the SU(3) limit, the CDD pole exists below the $\Lambda\Lambda$ threshold, and it appears above the $\Lambda\Lambda$ threshold at the physical point. By continuously varying the parameter $x$, we find that these poles are indeed connected with each other, as shown in Fig.~\ref{fig:CDD}. We thus conclude that the CDD pole in the SU(3) limit is the origin of the vanishing of the $\Lambda\Lambda$ amplitude near the $N\Xi$ threshold at the physical point.

\subsection{Extrapolation in the quark mass plane}

\begin{figure}[tbp]
    \centering
    \includegraphics[width=8cm]{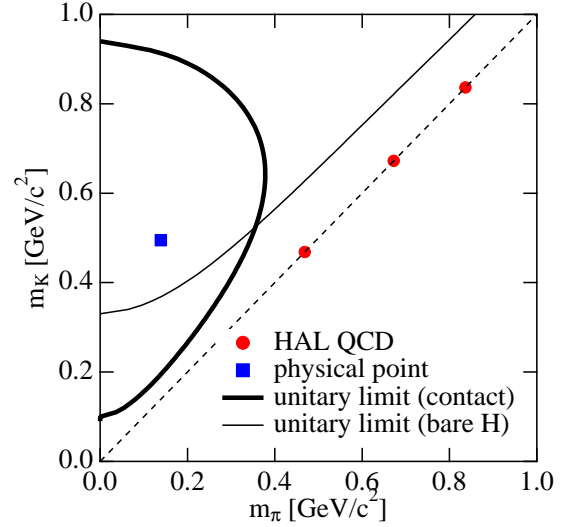}
    \caption{\label{fig:unitary_mNG}
    (Color online) Unitary limit in the $m_{\pi}$-$m_{K}$ plane in units of GeV. Thick (thin) line represents the result in the contact (bare H) model. 
Diagonal dotted line represents the SU(3) limit.
     }
\end{figure}%

\begin{figure}[tbp]
    \centering
    \includegraphics[width=8cm]{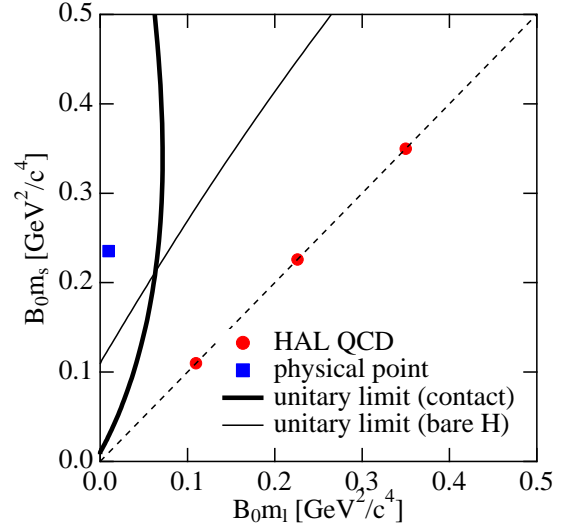}
    \caption{\label{fig:unitary_mq}
    (Color online) Unitary limit in the $m_{l}$-$m_{s}$ plane in units of GeV. Thick (thin) line represents the result in the contact (bare H) model. 
Diagonal dotted line represents the SU(3) limit.
    }
\end{figure}

We finally calculate the $\Lambda\Lambda$ scattering length with varying the quark masses $m_{l}$ and $m_{s}$. By identifying the unitary limit by the divergence of the $\Lambda\Lambda$ scattering length, we plot the unitary limit in the $m_{\pi}$-$m_{K}$ plane (Fig.~\ref{fig:unitary_mNG}) and in the $m_{l}$-$m_{s}$ plane (Fig.~\ref{fig:unitary_mq}). Qualitatively, it is a common feature that the unitary limit is realized between the physical point and the SU(3) limit. We however find that the location of the unitary limit in the quark mass plane highly depends on the model employed, in contrast to the previous results where the difference of the contact model and the bare H model is not very prominent. The difference in the extrapolation may have some significance in the charm sector where the $\Lambda_{c}\Lambda_{c}$ bound state is discussed~\cite{Meguro:2011nr}. To clarify the existence of the $\Lambda_{c}\Lambda_{c}$ bound state, we need to know the property at physical $m_{\pi}$ with a large $m_{K}\sim 1.87$ GeV/c$^{2}$, which is in the bound (unbound) region in the contact (bare H) model. More lattice data in the wide range of the quark mass plane will be helpful to pin down the exact location of the unitary limit.

{In Fig.~\ref{fig:unitary_mNG}, we also find the bound state at $(m_\pi,m_K)=(389,544)$ MeV, 
where the simulation of the NPLQCD collaboration
is performed~\cite{Beane:2010hg,Beane:2011iw}.
Although the obtained binding energy, 0.37 MeV (contact) and 0.046 MeV (bare H), is
smaller than 13.2 MeV reported by the NPLQCD,
it is qualitatively consistent with the NPLQCD results.
We note that the parameters of the EFT are not fitted to the NPLQCD results, and the baryon masses given by 
Eqs.~\eqref{eq:MN}-\eqref{eq:MXi} are slightly different from those in
the NPLQCD simulations.
}

The H-dibaryon in the chiral limit $m_{l}=m_{s}\to 0$ is of particular interest from the viewpoint of the Skyrmion~\cite{Balachandran:1983dj,Balachandran:1985fb}. However, the extrapolation in the present framework to the chiral limit should be performed with care, because the range of the NG boson exchange interaction is infinite in the chiral limit. Although the applicable energy region of the contact interaction model gradually reduces when we decrease the quark (NG boson) mass, the value of the scattering length at zero energy can be used to examine the existence of the bound state. The absence of the divergence of the scattering length on the SU(3) symmetric line in Figs.~\ref{fig:unitary_mNG} and \ref{fig:unitary_mq} indicates that the bound H-dibaryon found by the lattice QCD should remain bounded in the chiral limit. 

The realization of the unitary limit in Figs.~\ref{fig:unitary_mNG} and \ref{fig:unitary_mq} urges us to consider the BCS-BEC crossover~\cite{Zwerger} in the cold baryonic matter with strangeness under the variation of quark masses. In the present case, the bosonic bound state 
to be condensed in the BEC phase is the H-dibaryon. In this respect, we recall the discussion of the many-body system of (compact) H-dibaryons, the ``H-matter''~\cite{Tamagaki:1990mb,Sakai:1997dx}. If we regard the quark masses as controllable parameters (for instance, in lattice QCD simulation), we can tune them to realize a bound H-dibaryon in the two-body system, which eventually leads to the Bose-Einstein condensation in the many-body systems. It is an interesting possibility to consider the appearance of the ``H-matter'' in the unphysical quark mass region.

\section{Summary}\label{sec:summary}

We have studied the H-dibaryon in the $\Lambda\Lambda$ scattering from
the viewpoint of the quark mass dependence. The analysis is performed
with the effective field theory which universally describes the
low-energy scattering. Using the constraints by the lattice QCD data
{obtained by the HAL QCD collaboration}
to determine the quark mass dependence, we extrapolate the coupled-channel baryon-baryon scattering amplitude. 

The extrapolation to the physical point shows that the bound state found in the SU(3) limit disappears, and a weak resonance signal is found just below the $N\Xi$ threshold. Through the detailed study of the pole trajectory, we find that the bound state pole remains as a virtual state below the threshold, while a shadow pole of the bound state evolves to the resonance. We point out that the vanishing of the $\Lambda\Lambda$ scattering amplitude may cause the Ramsauer-Townsend effect, which originates in the CDD pole found in the SU(3) limit. It is shown that the coupled-channel effect is responsible for these nontrivial structures of the $\Lambda\Lambda$ scattering amplitude.

The extrapolation to the quark mass plane has various implications for the charm hadron sector and for the chiral limit. Among others, we show the existence of the unitary limit of the $\Lambda\Lambda$ scattering in between the physical point and the SU(3) limit. This opens the possibility to realize exotic phases of the finite density QCD by tuning of the quark masses.

\section*{Acknowledgments}

The authors thank Takashi Inoue, Kenji Sasaki, and Yoichi Ikeda for
useful discussions and providing the results of the scattering lengths
by the HAL QCD collaboration. This work is supported in part by JSPS
KAKENHI Grants No. 24740152 and No. 16K17694, by the Yukawa
International Program for Quark-Hadron Sciences (YIPQS), and by
INFN Fellowship Programme.


%

\end{document}